\documentclass[aps,prd,
preprint,
floatfix,nofootinbib,groupedaddress,showpacs]{revtex4}
\usepackage{graphicx}
\usepackage{amsmath}
\usepackage{latexsym}
\usepackage{here}
\usepackage{array}
\usepackage{dcolumn}
\usepackage{bm}
\usepackage{epsfig}
\usepackage{multirow}

\usepackage[dvips]{color}
\definecolor{Black}{named}{Black}
\definecolor{Red}{named}{Red}
\definecolor{Blue}{named}{Blue}

\newcommand{\Lumint}{{\cal L}_{\rm int}}

\def\lsim{\mathrel{\raise.3ex\hbox{$<$\kern-.75em\lower1ex\hbox{$\sim$}}}}

\def\epem{\ifmmode e^+e^-\else $e^+e^-$\fi}
\def\to{\rightarrow}

\def\mpl{\ifmmode \overline M_{Pl}\else $\bar M_{Pl}$\fi}
\def\beq{\begin{equation}}
\def\be{\begin{equation}}
\def\beqn{\begin{eqnarray}}
\def\ee{\end{equation}}
\def\eeq{\end{equation}}
\def\eeqn{\end{eqnarray}}

\begin{document}
\title{Precise determination of $Z$-$Z'$ mixing at the CERN LHC }

\author{
V. V. Andreev,$^{a,}$\footnote{quarks@gsu.by}\hspace{.4cm} 
P. Osland,$^{b,}$\footnote{per.osland@ift.uib.no}\hspace{.4cm}
A. A. Pankov$^{c,}$\footnote{pankov@ictp.it}\hspace{.4cm} }
\vspace{.5cm} \affiliation{
 $^{\rm a}$The F. Scorina Gomel State University, 246019 Gomel, Belarus\\
$^{\rm b}$Department of Physics and Technology, University of Bergen, Postboks 7803, N-5020  Bergen, Norway\\
 $^{\rm c}$The Abdus Salam ICTP Affiliated Centre, Technical University of Gomel, 246746 Gomel,
 Belarus
 }
\date{\today}
\begin{abstract}
We discuss the expected sensitivity to $Z^\prime$ boson effects in
the $W^\pm$ boson pair production process  at the Large Hadron
Collider (LHC). The results of a model-dependent analysis of
$Z^\prime$ boson effects are presented as constraints on the
$Z$-$Z'$ mixing angle $\phi$ and $Z^\prime$ boson mass.
The process $pp\to W^+W^- + X$ allows to place stringent
constraints on the $Z$-$Z^\prime$ mixing angle. Specifically, we
find that the present LHC bounds on the mixing angle are of the order
a few times $10^{-3}$, what is of the same order as those derived
from the electroweak data. These results were derived from
analysis of $W$-pair production at  $\sqrt{s}=8$ TeV and
integrated luminosity of 20 fb$^{-1}$. Further improvement on
the constraining of this mixing can be achieved from the analysis of
data on $WW\to l\nu l^\prime\nu^\prime$ ($l,l'=e$ or $\mu$) and
$WW\to l\nu jj$ final states collected at the LHC with nominal
energy and luminosity, 14 TeV and 100 fb$^{-1}$, and should
be $\phi\sim 10^{-4}-10^{-3}$.
\end{abstract}
\pacs{12.60.-i, 12.60.Cn, 14.70.Fm, 29.20.Ej}
\maketitle

\section{Introduction}
Nearly all electroweak and strong-interaction data are well described 
by the standard model (SM)~\cite{Beringer:2012}. However, there are many reasons why this is not
believed to be the ultimate theory. Grand Unified Theories (GUT), possibly
together with Supersymmetry, which allows a successful unification
of the three gauge coupling constants at the high scale, are among
the main candidates for new and richer physics. Many of
these GUTs, including superstring and left-right-symmetric models,
predict the existence of new neutral gauge bosons, which might be
light enough to be accessible at current and/or future
colliders~\cite{Hewett:1988xc,Langacker:2008yv,Leike:1998wr,Gulov:2010zq}.

The search for these $Z^{\prime}$ particles is an important aspect
of the experimental physics program of current and future
high-energy colliders. Present limits from direct production at
the LHC and  virtual effects at LEP, through interference or
mixing with the $Z$ boson, imply that new $Z^{\prime}$ bosons are
rather heavy and mix very little with the $Z$ boson. Depending on
the considered theoretical model, $Z^{\prime}$ masses of the
order of 2.5--3.0~TeV
\cite{Chatrchyan:2012oaa,atlas-dilepton,Aad:2014cka,CMS:2013qca} and $Z$-$Z^{\prime}$
mixing angles at the level of a few per mil are
excluded~\cite{Erler:2009jh,Andreev:2012zza,Aaltonen:2010ws}. 
The size of the mixing angle is strongly constrained by very high
precision $Z$-pole  experiments at LEP and the SLC \cite{ALEPH:2005ab}. They
contain measurements from the $Z$  line shape, from the leptonic branching
ratios normalized to the total hadronic $Z$ decay width and from leptonic
forward-backward asymmetries.
A $Z^{\prime}$ boson, if lighter than about 5 TeV, could be
discovered at the LHC \cite{Godfrey:2013eta,Dittmar:2003ir} 
with $\sqrt{s}=14$ TeV in the Drell-Yan
process
\begin{equation}\label{procDY}
pp \to Z' \to \ell^+ \ell^-+X
\end{equation}
with $\ell=e, \mu$. The future $e^+e^-$ International linear collider
(ILC) with high c.m.\ energies and longitudinally polarized beams
could indicate the existence of $Z^{\prime}$ bosons via its
interference effects in fermion pair production processes, with
masses up to about $6\times \sqrt{s}$ \cite{Rizzo:2006nw} while
$Z$-$Z'$ mixing will be constrained down to $\sim 10^{-4}-10^{-3}$ in
the process $e^+e^-\to W^+W^-$ \cite{Andreev:2012cj,Ananthanarayan:2010bt}.

After the discovery of a $Z^{\prime}$ boson at the
LHC via the process (\ref{procDY}), some diagnostics 
of its couplings and $Z$-$Z^{\prime}$ mixing
needs to be done in order to identify the correct theoretical
framework. In this paper we study the potential of the LHC to discover
$Z$-$Z'$ mixing effects in the process
\begin{equation}\label{procWW}
pp \to W^+ W^-+X
\end{equation}
and compare it with that expected at the ILC.

The $W^\pm$ boson pair production process (\ref{procWW}) is rather
important for studying the electroweak gauge symmetry at the LHC.
Properties of the weak gauge bosons are closely related to
electroweak symmetry breaking and the structure of the gauge
sector in general. In addition, the diboson decay modes of $Z'$
directly probe the gauge coupling strength between the new and the
standard-model gauge bosons. The coupling strength strongly
influences the decay branching ratios and the natural widths of
the new gauge bosons. Thus, detailed examination of the process
(\ref{procWW}) will both test the gauge sector of the SM
with the highest accuracy and throw light on New Physics (NP) that
may appear beyond the SM.

Direct searches for a heavy $WW$~resonance have been performed by
the CDF and D0 collaborations at the Tevatron. The D0
collaboration explored diboson resonant production using the
$\ell\nu \ell' \nu'$ and $\ell \nu j j$ final states~\cite{Abazov:2010dj}.
The CDF collaboration also searched for resonant $WW$ production
in the $e \nu j j$ final state, resulting in a lower limit on the
mass of an RS graviton, $Z'$ and $W'$ bosons~\cite{Aaltonen:2010ws}.

The direct $WW$ resonance search by the ATLAS Collaboration using
$l\nu l'\nu^\prime$ final-state events in 4.7 fb$^{-1}$ $pp$
collision data at the collider energy of 7 TeV set mass limits
on such resonances \cite{Aad:2013wxa,Aad:2012nev}. Also, the $l\nu
jj$ final state allows to reconstruct the invariant mass of the
system, under certain assumptions on the neutrino momentum from a $W$
boson decay.

In this note, we examine the feasibility of observing a
$Z^{\prime}$ boson in the $W^\pm$ pair production process at the
LHC, which in contrast to the Drell-Yan process (\ref{procDY}) is
not the principal discovery channel, but can help to understand
the origin of new gauge bosons. In the scenarios that we will consider in the following, the mechanism of $Z^\prime$ production and subsequent decay to $WW$ is directly proportional to the $Z$-$Z^\prime$ mixing.
Also, we show that the sensitivity of the $W^\pm$
pair production process in their pure leptonic decay channels to
the $Z$-$Z'$ mixing angle at the LHC with 8 TeV allows to place
limits on the mixing angle that are complementary to those derived
from the current electroweak data, whereas the increasing LHC
energy and time-integrated luminosity up to their planned values
allow to get corresponding limits that are competitive to
the current ones and to those expected from future ILC data. 

The paper is organized as follows. In Section~II, we briefly
review models involving additional $Z^\prime$ bosons and
emphasize the role of $Z$-$Z^\prime$ mixing in the process
(\ref{procWW}). In Section~III we give expressions for basic
observables, as well as formulae for helicity amplitudes of the
process under consideration. In Section~IV we discuss signals and backgrounds, both for the $ee,\mu\mu$ and for the $e\mu$ cases, and in Section~V we discuss achievable constraints on $Z^\prime$ models. Section~VI presents some concluding remarks.

\section{$Z'$ models}
There are many theoretical models which predict a $Z'$ with mass possibly in the
TeV range. Popular classes of models are represented by $E_6$-motivated models,
the Left-Right Symmetric Model (LR), the $Z'$ in an `alternative'
left-right scenario and the Sequential Standard Model (SSM), which
has a heavier boson with
couplings like those of the SM $Z$. Searching for $Z'$
in the above models has been widely studied in the literature
\cite{Langacker:2008yv,Leike:1998wr,Hewett:1988xc} and
applied at LEP2, the Tevatron and the LHC. 
For the notation we refer to \cite{Andreev:2012cj}, where also a brief description can be found.
The different models considered are: 
(i) Models related to the breaking of $E_{6}$, parametrized by a parameter $\beta$, familiar cases are the $Z^\prime_\chi$, $Z^\prime_\psi$,
$Z^\prime_\eta$ and $Z^\prime_I$ models;
(ii) Left-right models, originating from the breaking down of an
$SO(10)$ grand-unification symmetry, leading to a
$Z^\prime_{\rm LR}$;
(iii) The sequential $Z^\prime_\text{SSM}$, which has couplings
to fermions being the same as those of the SM $Z$.

The mass-squared matrix of
the $Z$ and $Z^{\prime}$ can have non-diagonal entries $\delta
M^2$, which are related to the vacuum expectation values of the
fields of an extended Higgs sector:
\begin{equation}\label{massmatrix}
M_{ZZ^\prime}^2 = \left(\begin{matrix} M_Z^2 & \delta M^2\\ \delta
M^2 & M_{Z^\prime}^2
\end{matrix}\right).
\end{equation}
Here, $Z$ and $Z^{\prime}$ denote the weak gauge boson eigenstates
of $SU(2)_L\times U(1)_Y$ and of the extra $U(1)'$, respectively.
The mass eigenstates, $Z_1$ and $Z_2$, diagonalizing the matrix
(\ref{massmatrix}), are then obtained by the rotation of the
fields $Z$ and $Z^\prime$:
\begin{subequations}
\label{Eq:Z12-couplings}
\begin{eqnarray}
&& Z_1 = Z\cos\phi + Z^\prime\sin\phi\;, \label{z1} \\
&& Z_2 = -Z\sin\phi + Z^\prime\cos\phi\;. \label{z2}
\end{eqnarray}
\end{subequations}
Here, the mixing angle $\phi$ is expressed in terms of masses as:
\begin{equation}
\label{phi} \tan^2\phi={\frac{M_Z^2-M_1^2}{M_2^2-M_Z^2}}\simeq
\frac{2 M_Z \Delta M}{M_2^2}\;,
\end{equation}
where $\Delta M=M_Z-M_1>0$, $M_Z$ being the mass of the $Z_1$ boson
in the absence of mixing, i.e., for $\phi=0$. Once we assume the
mass $M_1$ to be determined experimentally, the mixing depends on
two free parameters, which we identify as $\phi$ and $M_2$.

In general, such mixing effects reflect the underlying gauge
symmetry and/or the Higgs sector of the model. To a good
approximation, for $M_1\ll M_2$, in specific `minimal-Higgs
models',
\begin{equation}\label{phi0}
\phi\simeq -s^2_\mathrm{W}\
\frac{\sum_{i}\langle\Phi_i\rangle{}^2I^i_{3L}Q^{\prime}_i}
{\sum_{i}\langle\Phi_i\rangle^2(I^i_{3L})^2}
={\cal C}\ {\frac{\displaystyle M^2_1}{\displaystyle M^2_2}}.
\end{equation}
Here $\langle\Phi_i\rangle$ are the Higgs vacuum expectation
values spontaneously breaking the symmetry, and $Q^\prime_i$  are
their charges with respect to the additional $U(1)'$. In addition,
in these models the same Higgs multiplets are responsible for both
generation of mass $M_1$ and for the strength of the
$Z$-$Z^\prime$ mixing. Thus ${\cal C}$ is a model-dependent
constant. For example, in the case of $E_6$ superstring-inspired
models ${\cal C}$ can be expressed as \cite{UPR-0476T}
\begin{equation}\label{c}
{\cal C}=4s_\mathrm{W}\left(A-\frac{\sigma-1}{\sigma+1}B\right),
\end{equation}
where $s_\mathrm{W}$ is the sine of the electroweak  angle,
$\sigma$ is the ratio of vacuum expectation values squared,
and the constants $A$ and $B$ are determined by an angle
$\beta$ defining a direction in the extended gauge symmetry sector: 
$A=\cos\beta / 2\sqrt6$, $B=\sqrt {10}/12\sin\beta$.

An important property of the models under consideration is that
the gauge eigenstate $Z^{\prime}$ does not couple to the $W^+W^-$
pair since it is neutral under $SU(2)_L$. Therefore the process
(\ref{procWW}) is sensitive to a $Z^\prime$ only in the case of a
non-zero $Z$-$Z^{\prime}$ mixing.

From (\ref{Eq:Z12-couplings}), one obtains the vector and
axial-vector couplings of the $Z_1$ and $Z_2$ bosons to fermions:
\begin{subequations}
\begin{eqnarray}
&&\hspace{-10mm} v_{1f} = v_f\cos\phi + v_f^\prime
\sin\phi\;,\;a_{1f} = a_f \cos\phi +
a_f^\prime \sin\phi\;, \label{v1} \\
&&\hspace{-10mm} v_{2f} = v_f^\prime \cos\phi - v_f
\sin\phi\;,\;a_{2f} = a_f^\prime \cos\phi -a_f \sin\phi\;,
\label{v2}
\end{eqnarray}
\end{subequations}
with $(v_f,a_f)=(g^f_L\pm g^f_R)/2$, and $(v_f^\prime,a_f^\prime)$
similarly defined in terms of the $Z^\prime$ couplings. The
fermionic $Z^\prime$ couplings can be found, e.g.
in~\cite{Andreev:2012cj}.

Analogously, one obtains according to the remarks above:
\begin{subequations}
\begin{eqnarray}
&& g_{WWZ_1}=\cos\phi\;g_{WWZ}\;,  \\
&& g_{WWZ_2}=-\sin\phi\; g_{WWZ}\;,
\end{eqnarray}
\end{subequations}
where $g_{WWZ}=\cot\theta_W$.

In our analysis, we ignore kinetic mixing \cite{Holdom:1985ag}. Such mixing would introduce an additional parameter, and could modify the exclusion reach (see, for example \cite{Krauss:2012ku,Hirsch:2012kv}).

\section{Cross section}
The parton model cross section for the process (\ref{procWW}) from initial quark-antiquark states can be written as
\begin{equation}
 \frac{d\sigma_{q \bar q}}{dM\,dy\,dz} 
 = K \frac{2 M}{s}
\sum_q [f_{q|P_1}(\xi_1)f_{\bar q|P_2}(\xi_2) 
+ f_{\bar q|P_1}(\xi_1)f_{q|P_2}(\xi_2)]\, 
\frac{d\hat \sigma_{q \bar q}}{dz}\; . \label{dsigma}
\end{equation}
Here, $s$ is the proton-proton center-of-mass energy squared;
$z=\cos\theta$ with $\theta$ the $W^-$-boson--quark angle in the
$W^+W^-$ center-of-mass frame; $y$ is the diboson rapidity;
$f_{q|P_1}(\xi_{1},M)$ and $f_{\bar q|P_2}(\xi_{2},M)$ are parton
distribution functions in the protons $P_1$ and $P_2$,
respectively, with $\xi_{1,2}=(M/\sqrt s)\exp(\pm y)$ the parton
fractional momenta; finally, ${d\hat \sigma_{q \bar q}}/{dz}$ are
the partonic differential cross sections. In~(\ref{dsigma}), the
$K$ factor accounts for next-to-leading order QCD contributions 
\cite{Campbell:1999ah,Campbell:2011bn}
(for the invariant $WW$ mass-dependent cross section, see \cite{Agarwal:2010sp,Bai:2012zza}).
For simplicity, we will use as an
approximation a global flat value $K=1.2$ \cite{Agarwal:2010sp,Bai:2012zza} 
both for the SM and $Z'$
boson cases. For numerical computation, we use CTEQ-6L1 parton
distributions \cite{Pumplin:2002vw}. 
Since our estimates will be at the Born level, the factorisation scale $\mu_{\rm F}$
enters solely through the parton distribution functions, as the
parton-level cross section at this order does not depend on
$\mu_{\rm F}$.
As regards the scale dependence of the parton distributions we choose for the factorization scale the $WW$ invariant mass, i.e., $\mu_{\rm F}^2=M^2=\hat{s}$, with $\hat{s}=\xi_1 \,\xi_2\,s$  the parton subprocess c.m.\ energy squared.
We have checked that the obtained constraints presented in the following are not significantly modified when $\mu_{\rm F}$ is varied in the interval $\mu_{\rm F}/2$ to $2\mu_{\rm F}.$

Taking into account the experimental  rapidity cut relevant to the
LHC experiments, ($Y_{\rm{cut}}=2.5$), one should carry out the
integration over the phase space in (\ref{dsigma}) determined as
\cite{Nuss1997,Osland:2008sy}:
\begin{equation}
\left|y \right|\leq Y=\min\left[\ln(\sqrt{s}/M),
Y_{\rm{cut}}\right]=\ln(\sqrt{s}/M), \label{parton2}
\end{equation}
where we make use of the fact that we do not consider low masses, 
$\ln(\sqrt{s}/M)<Y_\text{cut}$.
This leads to a cut in the production angle
\begin{equation}
\left|z\right|\leq
z_{\mathrm{cut}}=\min\left[\tanh(Y_{\rm{cut}}-|y|)/\beta_W,1\right]\;, 
\label{parton3}
\end{equation}
where $\beta_W=\sqrt{1-4M_W^2/\hat{s}}$ and $M_W$ is the $W$ boson
mass.

The resonant $Z'$ production cross section of process (\ref{procWW}) needed in order to estimate the expected number of $Z'$ events, can be derived from (\ref{dsigma}) by integrating
its right-hand-side  over $z$, the rapidity of the $W^\pm$-pair $y$ and
invariant mass $M$ around the resonance peak $(M_R-\Delta M/2,$ $
M_R+\Delta M/2)$:
\begin{equation}
\sigma{(pp\to W^+W^- + X)}  
=\int_{M_{R}-\Delta M/2}^{M_{R}+\Delta M/2}d M
\int_{-Y}^{Y}d y \int_{-z_{\text{cut}}}^{z_\text{cut}}d
z\frac{d\sigma_{q \bar q}}{d M\, d y\, d z}\;. \label{TotCr}
\end{equation}
We adopt the parametrization of the experimental mass resolution
$\Delta M$ in reconstructing the diboson invariant mass of the
$W^+W^-$ system, $\Delta M$ vs.\ $M$,  as proposed in
Ref.~\cite{Atlas}. (After integration over $y$, interference effects vanish.)

\begin{figure}[htb] \vspace{3mm}
\begin{center}
{\includegraphics[angle=90, scale=0.55]{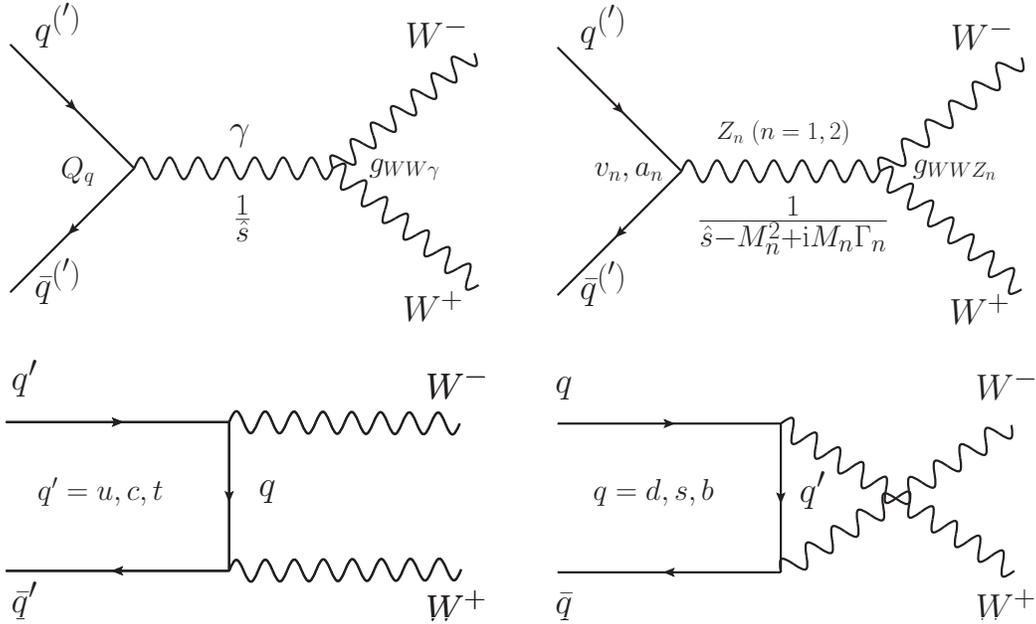}}
\end{center}
\caption{Feynman diagrams of the $q\bar{q}$ $(q^\prime\bar{q}^\prime)
\to W^+W^-$ process  within the framework of the extended gauge models.}\label{fig1}
\end{figure}

The parton level $W^\pm$ boson pair production can be described, within the gauge models discussed here, by the subprocesses
\begin{equation}
q\bar{q} \to {\gamma,Z_1,Z_2} \to W^+W^-,\label{parton}
\end{equation}
as well as $t$- and $u$-channel Feynman diagrams displayed in
Fig.~\ref{fig1}.

The differential (unpolarized) cross section of process~(\ref{parton})
can be written as:
\begin{equation}
\frac{d{\hat\sigma}_{q \bar q}}{d z
}=\frac{1}{N_C}\,\frac{\beta_W}{32\pi
\hat{s}}\sum_{\lambda,\lambda',\tau,\tau'}\vert
F_{\lambda\lambda'\tau\tau'}(\hat{s},\theta)\vert^2\;.\label{xsection}
\end{equation}
Here, $N_C$ is the number of quark colors; $\lambda=-\lambda'= \pm
1/2$ are the  quark helicities; the helicities of the $W^-$ and
$W^+$ are denoted by $\tau,\tau'=\pm 1,0$.   The helicity
amplitudes $F_{\lambda\lambda'\tau\tau'}(\hat{s},\theta)$ are
summarized in Table~\ref{amplit} that  reproduces the SM
expectations if one ignores the effects of the $Z$-$Z'$ mixing
\cite{Nuss1997,Gounaris:1992kp,Gounaris:2013sva}. There $\hat{s}$, $\hat{t}$, $\hat{u}$ are the
Mandelstam variables defined as $\hat{t}=M_W^2-\hat{s}(1 \hm
-\beta_W z)/2$, $\hat{u} =M_W^2-\hat{s}(1 + \beta_W z)/2 $;
$\Gamma_{1,2}$ are $Z_{1,2}$ boson decay widths;
$g^\lambda_{1,f}=v_{1,f}-2a_{1,f}\lambda$,
$g^\lambda_{2,f}=v_{2,f}-2a_{2,f}\lambda$; and $\gamma_{W} =
\sqrt{\hat{s}}/2M_W$. 
In the $t$- and $u$-channel exchanges of Fig.~\ref{fig1} we account for the initial 
$q = u,d,s,c$, only the CKM favoured quarks in the approximation of unity relevant matrix element.

\begin{table}[h t p b]
\centering \caption{ Helicity amplitudes \cite{Nuss1997} of $q\bar{q} \to W^+W^-$. To obtain the amplitude
$F_{\lambda\lambda'\tau\tau'}(\hat{s},\theta)$ for a definite quark
helicity $\lambda=-\lambda'=\pm 1/2$ and fixed helicities $\tau
(W^-)$ and $\tau^\prime (W^+)$ of the final-state system, it is
necessary to multiply each element of the respective column by the
common factor in its upper part, to multiply successively the
elements obtained in this way by the respective elements in the
first column, and to perform thereupon summation over all
intermediate states.} \label{amplit} \vspace{3mm}
\begin{tabular}{|>{\small}c|>{\small}c|>{\small}c|>{\small}c|}

\hline {} & \rule{0mm}{5mm} {helicity of $W^\pm$} &
$\tau=\tau'=\pm 1$ & $\tau=-\tau'=\pm 1$ \\ [1mm]

\hline channel & \rule{0mm}{5mm} $W^-_T W^+_T$ & $-e^2 \hat{s}
\lambda \sin\theta/2$ & $-e^2 \hat{s} \lambda \sin\theta/2$ \\
[1.2mm]

\hline t & \rule{0mm}{5mm} $\frac{2\lambda -1}{4 \hat{t}
s^2_\mathrm{W}}$ & $\cos\theta - \beta_{W}$ & $-\cos\theta
-2\tau\lambda$ \\ [1.2mm]

\hline u & \rule{0mm}{5mm} $\frac{2\lambda -1}{4 \hat{u}
s^2_\mathrm{W}}$ & $\cos\theta + \beta_{W}$ & $-\cos\theta
-2\tau\lambda$ \\ [1.2mm]

\hline s & \rule{0mm}{5mm} $\frac{2Q_f}{\hat{s}} + g^\lambda_{1,f}\frac{2 g_{WWZ_1}}{\hat{s}-M_1^2+i M_1\Gamma_1}$ & $-\beta_{W}$ & $0$ \\
{} & $+g^\lambda_{2,f}\frac{2 g_{WWZ_2}}{\hat{s}-M_2^2+i
M_2\Gamma_2}$ & {} & {} \\ [1.5mm]


\hline \hline {} & \rule{0mm}{5mm} {helicity of $W^\pm$} & $\tau=\tau'=0$ & {} \\
[1mm]

\hline channel& \rule{0mm}{5mm} $W^-_L W^+_L$ & $-e^2 \hat{s}
\lambda \sin\theta/2$ & {}
\\ [1.2mm]

\hline t & \rule{0mm}{5mm} $\frac{2\lambda -1}{4 \hat{t}
s^2_\mathrm{W}}$ &
$2\gamma_W^2\left(\cos\theta - \beta_W (1+\frac{1}{2\gamma_W^2}) \right)$ & {} \\
[1.2mm]

\hline u & \rule{0mm}{5mm} $\frac{2\lambda -1}{4 \hat{u}
s^2_\mathrm{W}}$ & $2\gamma_W^2\left(\cos\theta + \beta_W
(1+\frac{1}{2\gamma_W^2}) \right)$ & {}
\\ [1.2mm]

\hline s & \rule{0mm}{5mm} $\frac{2Q_f}{\hat{s}} + g^\lambda_{1,f}\frac{2 g_{WWZ_1}}{\hat{s}-M_1^2+i M_1\Gamma_1}$ & $-\beta_{W}(1+2\gamma_{W}^{2})$ & {} \\
{} & $+g^\lambda_{2,f}\frac{2 g_{WWZ_2}}{\hat{s}-M_2^2+i
M_2\Gamma_2}$ & {} & {} \\ [1.5mm]


\hline \hline {} & \rule{0mm}{5mm} {helicity of $W^\pm$} &
$\tau=0, \tau'=\pm 1$ & $\tau=\pm 1,\tau'=0$
\\ [1mm]

\hline channel & \rule{0mm}{5mm} $W^-_L W^+_T + W^-_T W^+_L$ &
$\frac{-e^2 \hat{s} \lambda}{2\sqrt{2}}
\left(\tau'\cos\theta-2\lambda\right)$ & $\frac{-e^2 \hat{s}
\lambda}{2\sqrt{2}} \left(\tau\cos\theta+2\lambda\right)$ \\
[1.2mm]

\hline t & \rule{0mm}{5mm} $\frac{2\lambda -1}{4 \hat{t}
s^2_\mathrm{W}}$
& $\gamma_{W}\left[\cos\theta(1+\beta_{W}^2)-2\beta_{W}\right]$ & $-\gamma_{W}\left[\cos\theta(1+\beta_{W}^2)-2\beta_{W}\right]$ \\
{} & {} &
$-\frac{\tau'\sin^2\theta}{\gamma_{W}\left(\tau'\cos\theta-2\lambda\right)}$
& $+\frac{\tau\sin^2\theta}{\gamma_{W}\left(\tau\cos\theta+2\lambda\right)}$ \\
[1.0mm]

\hline u & \rule{0mm}{5mm} $\frac{2\lambda -1}{4 \hat{u}
s^2_\mathrm{W}}$
& $\gamma_{W}\left[\cos\theta(1+\beta_{W}^2)+2\beta_{W}\right]$ & $-\gamma_{W}\left[\cos\theta(1+\beta_{W}^2)+2\beta_{W}\right]$ \\
{} & {} &
$-\frac{\tau'\sin^2\theta}{\gamma_{W}\left(\tau'\cos\theta-2\lambda\right)}$
&
$+\frac{\tau\sin^2\theta}{\gamma_{W}\left(\tau\cos\theta+2\lambda\right)}$
\\ [1.0mm]

\hline s & \rule{0mm}{5mm} $\frac{2Q_f}{\hat{s}} +
g^\lambda_{1,f}\frac{2
g_{WWZ_1}}{\hat{s}-M_1^2+i M_1\Gamma_1}$ & $-2 \beta_{W} \gamma_{W}$ & $2 \beta_{W} \gamma_{W}$ \\
{} & $+g^\lambda_{2,f}\frac{2 g_{WWZ_2}}{\hat{s}-M_2^2+i
M_2\Gamma_2}$ & {} & {} \\ [1mm] \hline
\end{tabular}
\end{table}

In evaluation of the total width $\Gamma_2$ of the $Z_2$ boson we take
into account its decay channels into fermions (quarks and
leptons) and $W^\pm$ boson pair
\cite{Pankov:1992cy}:
\begin{equation}\label{gamma2}
\Gamma_{2} = \sum_f \Gamma_{2}^{ff} + \Gamma_{2}^{WW}\;.
\end{equation}
Further contributions of decays involving Higgs and/or gauge
bosons and supersymmetric partners (including sfermions),
which are not accounted for in (\ref{gamma2}), could increase
$\Gamma_2$ by a model-dependent amount typically as large as 50\%
\cite{Pankov:1992cy}. For a discussion of width effects, see \cite{Accomando:2013sfa}.

The fermion contribution, $\sum_f \Gamma_{2}^{ff}$,  depends on the number $n_g$ of
generations of heavy exotic fermions which can contribute to $Z_2$
decay without phase space suppression (we can assume that the
three known generations do contribute). This number is model
dependent too, and brings a phenomenological uncertainty. For the
range of $M_2$ values assumed here, of the order of a few TeV, the
dependence of $\Gamma_2$ on $\phi$ induced by $\sum_f
\Gamma_{2}^{ff}$  and by $\Gamma_{2}^{WW}$ is unimportant. For
definiteness the $Z_2$ width $\Gamma_2$ is assumed to scale with
the $Z_2$ mass $\Gamma_2=(M_2/M_1)\Gamma_1\approx 0.03\,M_2$. This
scaling is what would be expected for the reference model SSM
\cite{Benchekroun:2001je}. Choosing this scaling is a conservative
assumption since in $E_6$ models, the $Z_2$ width would be
substantially narrower than this (see Table~\ref{tab2}).

\begin{table}[htb]
\caption{Ratio $\Gamma_{2}/M_{2}$ for the $\chi, \psi, \eta$ and SSM
models.}
\begin{center}
\begin{tabular}{|c|c|}
\hline $Z^\prime$  & $\Gamma_{2}/M_{2}$ [\%] \\
\hline $\chi$ & 1.2 \\
\hline $\psi$ & 0.5 \\
\hline $\eta$ & 0.6 \\
\hline SSM & 3.0 \\
\hline
\end{tabular}
\end{center}
\label{tab2}
\end{table}

The differential cross section for the processes $q\bar{q}\to Z'\to
W^+W^-$, averaged over quark colors, can easily be obtained from
Eq.~(\ref{xsection}) and written as \cite{delAguila:1987af}
\begin{align}
 \frac{d\hat{\sigma}^{Z'}_{q \bar q}}{d \cos\theta}
 &= \frac{1}{3}\,\frac{\pi\alpha^2 \cot^2\theta_W}{16 }
\beta_W^3\left(v_{2,f}^2 + a_{2,f}^2\right)\sin^2\phi \,
\frac{\hat{s}}
{\left(\hat{s} - M_{2}^2\right)^2 + M_{2}^2\Gamma_{2}^2}  \nonumber \\
&  \times\left(\frac{\hat{s}^2}{M_W^4} \sin^2\theta +
4\frac{\hat{s}}{M_W^2}(4-\sin^2\theta)+12\sin^2\theta\right).
\label{xsection2}
\end{align}
The resonant production cross section of process (\ref{procWW}) at
hadronic level can be derived from Eqs.~(\ref{dsigma}) and
(\ref{xsection2}).

It is important to notice that the dominant term in
Eq.~(\ref{xsection2}), for $M^2\gg M_W^2$, is proportional to
$(M/M_W)^4\cdot\sin^2\theta$ and corresponds to the production of
longitudinally polarized $W$'s, $Z'\to W^+_LW^-_L$. This 
increasing (with the parton sub-energy squared $\hat s$)  behavior of the cross section in the $Z^\prime$ scenarios considered in Table~II,  would, in turn, result in a corresponding enhanced sensitivity to $Z$-$Z'$ mixing at high $M$.

\begin{figure}[htb]
\begin{center}
{\includegraphics[angle=0, scale=1.2]{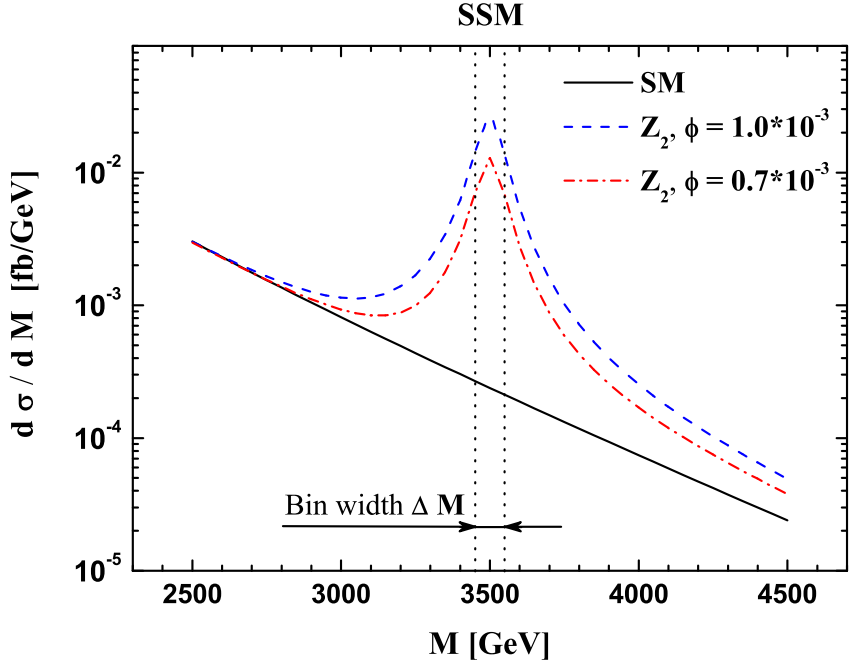}}
\end{center}
\caption{Invariant mass  distribution of $W^\pm$ pairs in
$pp\to W^+W^-+ X$ in the SM (solid curve) and for the $Z^\prime_{\rm
SSM}$ model ($M_{Z^\prime}=3.5~\text{TeV}$) with $Z$-$Z'$ mixing angle of $\phi=10^{-3}$ (dashed
line) and $\phi=0.7\cdot 10^{-3}$ (dash-dotted line) at the LHC
with $\sqrt{s}=14$~TeV.}
\label{fig2}
\end{figure}

For illustrative purposes, the invariant mass distribution of
$W^\pm$ pairs in the process $pp\to W^+W^+ + X$ in the SM (solid
black curve) and for the $Z^\prime_{\rm SSM}$ model at two vaues of the  $Z$-$Z'$
mixing angle at the LHC with
$\sqrt{s}=14$ TeV  is shown in Fig.~\ref{fig2}. 
The $W^\pm$-pair invariant mass
distribution ($d\sigma/dM$) is calculated with the same
parton distribution functions and event selection
criterion as those used in Ref.~\cite{Agarwal:2010sn}.
Also, the bin size
$\Delta M$ of the diboson invariant mass is depicted  for
comparison with the $Z^\prime$ width. 
For numerical computations, we take $\Delta M=0.03M$.
The $W$ bosons are kept on-shell and their subsequent decays are not included 
in the crosss sections represented in Fig.~\ref{fig2}. Here, we assumed that the
invariant mass distribution of the cross section can be
reconstructed from the decay products of the $W^+W^-$.
Fig.~\ref{fig2} shows that  at the LHC with integrated luminosity
${\cal L}_{\rm int}=100~{\rm{fb}}^{-1}$ the expected number of
$W^+W^-$ background events within a mass bin $\Delta M$ is of the order
of a few events while the resonant yield  at $\phi=10^{-3}$ is
$N_{Z'}\sim 100$.

\section{Signal and backgrounds}
In this section we focus on the $WW$ production via intermediate $Z'$ and subsequent 
purely leptonic decay of on-shell $W'$s, that will be probed experimentally at LHC, namely:
\begin{equation}
pp\to WW+X\to l\nu l^\prime\nu^\prime +X\label{WWll} \quad
(l,l'=e \text{ or }\mu),
\end{equation}
and, following the analysis given in \cite{Alves:2009aa,Eboli:2011bq}, we briefly introduce 
the main backgrounds and possible cuts to enhance the $Z'$ signal to background ratio.
The $WW\to \tau\nu l\nu$ and $WW\to\tau\nu\tau\nu$ processes with
$\tau$ leptons decaying into electrons and muons with additional
neutrinos are also included. Three final states are considered,
based on the lepton flavor, namely $ee$, $\mu\mu$, and $e\mu$
\cite{Aad:2012nev}. The branching fraction of the decay channels $WW$ into $e^+e^-$,
$e^+\mu^-$ and $\mu^+\mu^-$ pairs can be found in
Table~\ref{branch}.

\begin{table}[htb]
\caption{Branching fractions of the $WW\to l\nu
l^\prime\nu^\prime$ ($l,l'=e$ or $\mu$) decay channels
\cite{Cooke_2008}. }
\begin{center}
\begin{tabular}{|c|c|}
\hline process & Br [\%] \\
\hline $WW\to ee$ & 1.16 \\
$WW\to \tau e /\tau\tau\to ee$ & 0.47 \\
\hline $WW\to e\mu$ & 2.27 \\
$WW\to\tau l\to e\mu$ & 0.92 \\
\hline $WW\to \mu\mu$ & 1.12 \\
 $WW \to\tau\mu /\tau\tau\to\mu\mu$ & 0.45\\ \hline
\end{tabular}
\end{center}
\label{branch}
\end{table}

The presence of (at least) two neutrinos in the final state makes almost
impossible the complete reconstruction of the $WW$ invariant mass,
so that in any case the $Z'$ peak would be broadened. (For the pure leptonic channel
discussed below, the actual final-state width will be broader \cite{Alves:2009aa},
however this is not the case for the semileptonic channel.)
For an attempt to reconstruct the $M_{WW}$ distribution by means of estimating the momenta of escaping neutrinos, see, for example, Ref.~\cite{Eboli:2011bq}. Alternatively, a possible kinematical variable to characterize both the $Z'$ signal and the background should be the transverse $WW$ mass $M_T^{WW}$ (a $Z^\prime$ would lead to an excess of events at $M_T^{WW}>M_{Z^\prime}/2$), which has the advantage that only (measurable) transverse momenta are involved \cite{Alves:2009aa}. Both methods seem to lead to similar results, namely, the distributions of events are dominated by characteristic $Z'$ bumps over the backgrounds.

\subsection{Different-flavor leptons}
In the case of different-flavor leptons ($e\mu$), the $Z'$ signal in the
process (\ref{WWll}) possesses SM backgrounds coming from the
production of $W^+W^-$ pairs with its subsequent leptonic decay.

In order to perform effectively the detection and isolation of the
final leptons with opposite charges, paralleling Ref.~\cite{Alves:2009aa} we apply:
\begin{equation}
\vert\eta_l\vert<2.5,\,\, \Delta R_{ll}>0.4,\,\, p_T^l> 50\, {\rm
GeV},\label{cuts1}
\end{equation}
where $\Delta R_{ll}=\sqrt{(\Delta\eta)^2+(\Delta\phi)^2}$ parametrizes the separation in rapidity $\eta$ and azimuthal angle $\phi$.

Apart from the SM mechanism, another potentially sizable source of background arises  from $t\bar{t}$ pair production where the top quarks decay to $b+W$, leading to $b$-jets. This
background can be efficiently reduced by vetoing the presence of
additional jets with
\begin{equation}
    | \eta_j | < 3   \quad \hbox{ and } \quad p_T^j > 20~\hbox{ GeV.}
\label{eq:veto}
\end{equation}

However, it is necessary to account for the possible appearance of an additional jet in the signal event sample, originating either from QCD gluon radiation or from the pile-up of $pp$ interactions caused by the high luminosity. Accordingly, one can introduce probabilities for survival to the central jet veto (\ref{eq:veto}) of QCD and electroweak events, and the following values are found \cite{Eboli:2011bq}:
\begin{equation}
P_{\rm surv}^{\rm EW}= 0.56, \qquad P_{\rm
surv}^{\rm QCD}=0.23. \label{eq:vsp}
\end{equation}
The above constraints will be included in the statistical analysis
carried out in the next section.
Of course, an even stronger background suppression, relative to the $Z'$ signal, might be obtained by imposing the reconstructed $WW$ mass to coincide within a width with the [possibly determined fron DY] $Z'$ mass.

\subsection{Same-flavor leptons}
For same-flavor leptons ($ee$ or $\mu\mu$) there are additional
backgrounds originating from Drell-Yan lepton pair production, and
from the $ZZ$ production with one $Z$ decaying into charged leptons
and the other decaying invisibly or with both $Z$s decaying into
charged leptons, two of which escape undetected. In this case, two
extra cuts should be imposed in order to suppress the Drell--Yan
pair production and $ZZ$ background, namely
\begin{equation}
     E^\text{miss}_T > 50~\text{GeV}, \quad
      m_{\ell^+ \ell^-} > 100~\text{GeV,}
\label{etcut}
\end{equation}
respectively. As was concluded in
\cite{Alves:2009aa,Eboli:2011bq}, after applying the cuts, the
electroweak background originating from $t\bar{t}$ pair production
at high $Z'$ masses becomes negligible with respect 
to the irreducible background induced from $W^+W^-$ pair production via the SM.

\section{Constraints on $Z'$}
In our analysis, we denote by $N_{\rm SM}$ and $N_{Z'}$ the numbers
of `background' and `signal' events, and we adopt the criterion
$N_{Z'}=2{\sqrt{N_{\rm SM}}}$ or 3 events, whichever is larger, as
the minimum signal for reach at the 95\% C.L.
\cite{Leike:1998wr}. Here, the $Z'$ signal can be determined as
\begin{equation} \label{Eq:NZprime}
N_{Z'}={\Lumint}\times \sigma^{Z'} \times P_{\rm surv}^{\rm EW}
\times A\times\epsilon^{\ell},
\end{equation}
with
\begin{equation}
\sigma^{Z'}=\sigma(pp\to
Z')\times \text{Br}(Z'\to W^+W^-\to l \nu l'\nu^\prime).
\end{equation}
In Eq.~(\ref{Eq:NZprime}), $\Lumint$ is the time-integrated luminosity, and
$A\times\epsilon^{\ell}$ is the product of the overall acceptance times
the lepton detection and reconstruction efficiencies where $A$ represents
the kinematic and geometric acceptance from the total phase space to the
fiducial phase space governed by Eqs.~(\ref{parton2}) and (\ref{parton3}),
while $\epsilon^{\ell}$ represents detector effects such as lepton trigger
and identification efficiencies. The overall acceptance times the lepton
efficiency is $W^\pm$ invariant mass dependent and, for simplicity, we
take that to be 0.5.
The SM background reads:
\begin{equation}
N_{\rm SM}={\Lumint}\times \left(
\sigma^{\rm EW}_{\rm SM}\times
  P_{\rm surv}^{\rm EW} +\sigma^{t\bar t}_{\rm SM}\times P_{\rm
    surv}^{\rm QCD} \right) \times A\times\epsilon^{\ell} \approx
 {\Lumint}\times  \sigma^{\rm EW}_{\rm SM}\times
  P_{\rm surv}^{\rm EW}  \times A\times\epsilon^{\ell},
\end{equation}
where $\sigma^{\rm EW}_{\rm SM}$ is determined by Eqs.~(\ref{TotCr}) and
(\ref{xsection})  taking into account solely the SM contribution.
Also, in the latter expression for $N_{\rm SM}$ we take into account
that for heavy $M_{Z'}$, $\sigma^{\rm EW}_{\rm SM}\gg\sigma^{t\bar
t}_{\rm SM} $ as was shown in \cite{Alves:2009aa}.

One should notice that the latter estimation of the SM background,
$\sigma^{\rm EW}_{\rm SM}$, is consistent with what is obtained by
using the so-called MAOS method to reconstruct the $WW$ invariant mass described
in \cite{Eboli:2011bq}. However, we numerically find that at high
$Z'$ mass, $M_{Z'}>3$~TeV, the SM background becomes so low
that the criterion $N_{Z'}=3$ can be applied in obtaining
constraints on $Z'$ parameters.

\begin{figure}[hbt] \vspace{-10mm}
\begin{center}
{\includegraphics[angle=0, scale=1.05]{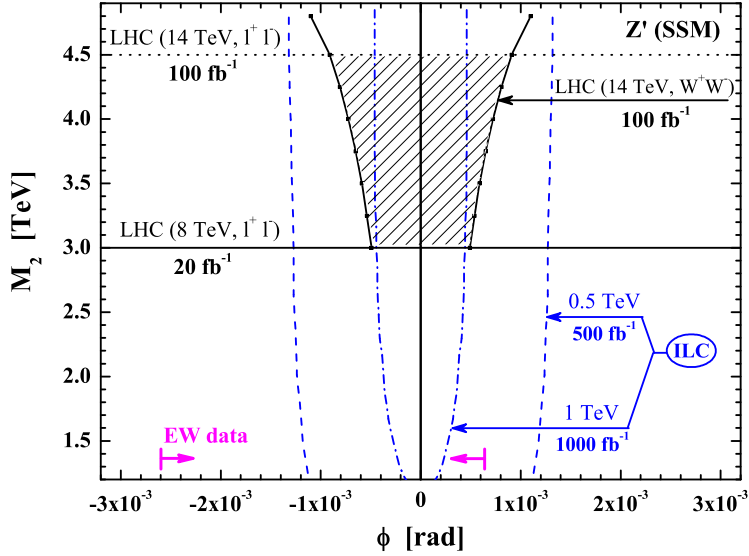}}
\end{center}
\vspace{-10mm}
\caption{Reach (at 95 \%C.L.) on $Z$-$Z'$ mixing and
$M_2$ mass for $Z'_{\rm SSM}$ obtained from the inclusive process $pp\to WW\to
l\nu l^\prime\nu^\prime$ ($l,l'=e$ or $\mu$) at the LHC (solid lines).
The allowed domain in $\phi$ and $M_2$ is the hatched one.
Current  limits on $M_2$ for $Z'_{\rm SSM}$ derived from the Drell--Yan
($l^+l^-$) process at the LHC (8 TeV) (horizontal solid line) as
well as `typical' mass limits expected at the LHC (14 TeV)
(horizontal dotted line) are shown. Limits on the $Z$-$Z'$
mixing angle from electroweak precision data
are displayed, and those expected from  $W^\pm$ pair production at
the ILC with polarized beams.}\label{fig3}
\end{figure}

\subsection{Leptonic $WW$ decays}

\begin{figure}[htb] \vspace{-6mm}
\begin{center}
{\includegraphics[angle=0, scale=1.1]{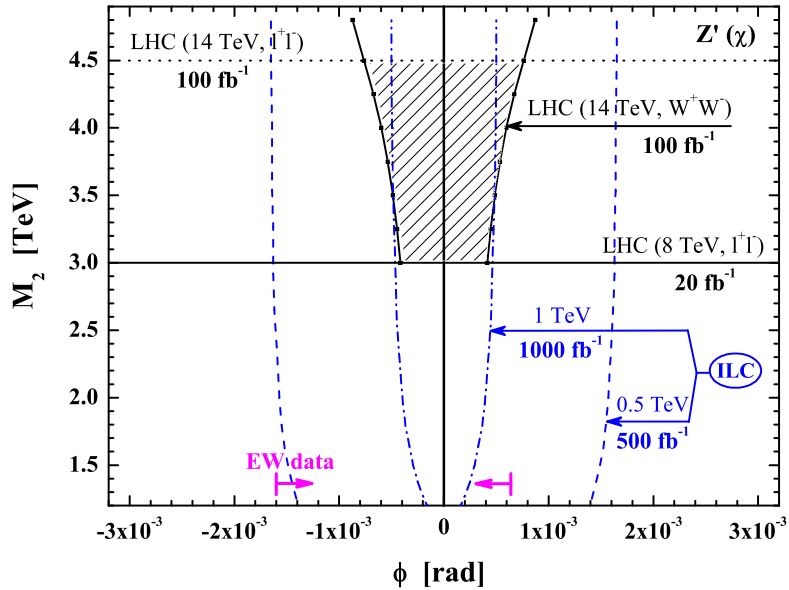}}
\end{center}
\vspace{-6mm}
\caption{ Same as in Fig.~3 but for $Z'_\chi$.
 }\label{fig4}
\end{figure}

\begin{figure}[htb] \vspace{-6mm}
\begin{center}
{\includegraphics[angle=0, scale=1.1]{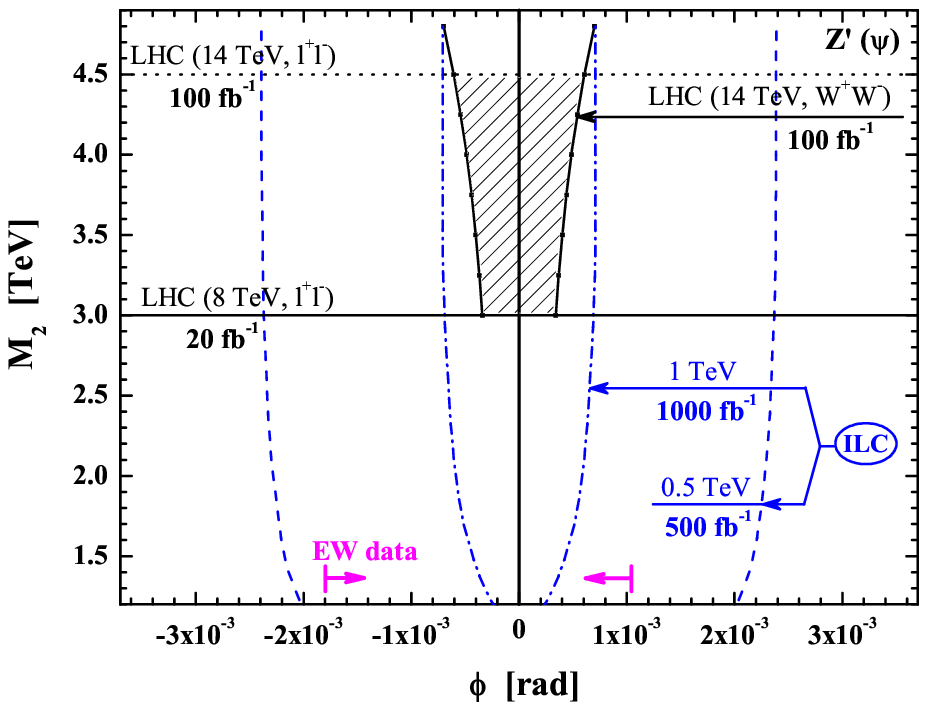}}
\end{center}
\vspace{-6mm}
\caption{ Same as in Fig.~3 but for $Z'_\psi$.
 }\label{fig5}
\end{figure}

We depict in Figs.~\ref{fig3}--\ref{fig5} the region in parameter space
to which the LHC will be able to constrain
$Z$-$Z^\prime$ mixing for an integrated luminosity of
100 fb$^{-1}$.

In particular, the discovery reach on the $Z$-$Z'$ mixing and $M_2$ mass for
$Z'_{\rm SSM}$ obtained from the process $pp\to WW +X\to l\nu
l^\prime\nu^\prime + X$ ($l,l'=e$ or $\mu$) at the LHC with
$\sqrt{s}=14$ TeV and $\Lumint=100$ fb$^{-1}$ are depicted by the two
solid lines. The form of these bounds is governed by the
criterion of $N_{Z'}=3$ and the quadratic dependence of the resonant
cross section, Eq.~(\ref{xsection2}), on the $Z$-$Z'$ mixing angle.
Also, current limits on $M_2$ for $Z'_{\rm SSM}$ derived from
the Drell--Yan ($l^+l^-$) process at the LHC (8 TeV) (horizontal solid
line) as well as those expected from the future experiments at the
LHC with 14 TeV (horizontal dotted line) are shown. The combined
allowed area in the ($\phi, M_2$) plane obtained from the Drell--Yan
and $W^\pm$ pair production processes is shown as a hatched
region. In addition, present limits on the $Z$-$Z'$ mixing angle
obtained from electroweak precision data analysis
\cite{Erler:2009jh} labelled as `EW data' are displayed (these have a weak mass dependence which we have not attempted to draw). For
comparison, the corresponding limits obtained from  $W^\pm$ pair
production at the ILC with polarized beams and for two options of
energy and time-integrated luminosity (0.5 (1) TeV and 0.5 (1)
ab$^{-1}$, respectively) are also presented \cite{Andreev:2012cj}.
Figs.~\ref{fig3}--\ref{fig5} show that the LHC is able to not only
significantly improve the current limits on the $Z$-$Z'$ mixing angle, but
in several cases, also allow more stringent bounds than those expected from future experiments
on the $WW$ channel at the electron--positron collider ILC \cite{Andreev:2012zza}.

\begin{figure}[htb] \vspace{-6mm}
\begin{center}
{\includegraphics[angle=0, scale=1.1]{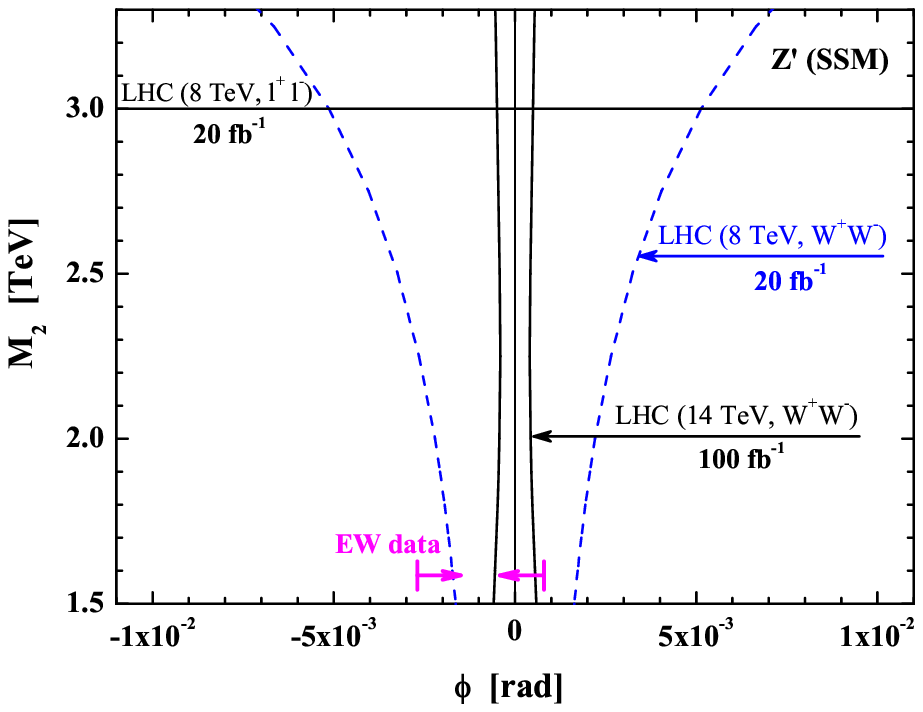}}
\end{center}
\vspace{-6mm}
\caption{Reach (at 95 \%C.L.) on $Z$-$Z'$ mixing and
$M_2$ mass for $Z'_{\rm SSM}$ obtained from the inclusive process $pp\to WW\to
l\nu l^\prime\nu^\prime$ at the LHC with $\sqrt{s}=8$ TeV and
$\Lumint=20$ fb$^{-1}$ (dashed curves) and $\sqrt{s}=14$ TeV and
$\Lumint=100$ fb$^{-1}$ (solid curves). Also shown are current limits on
$M_2$ derived from the Drell--Yan process at the LHC (8 TeV) denoted by
the label $l^+l^-$ and constraints from electroweak precision data. Note that the scale is different from that of Figs.~\ref{fig3}--\ref{fig5}.}
\label{fig6}
\end{figure}

In Fig.~\ref{fig6} we return to the $Z'_{\rm SSM}$ and show the sensitivity reach (at 95 \%C.L.) on
the $Z$-$Z'$ mixing and $M_2$ mass obtained from the $W$
pair production process at the LHC with $\sqrt{s}=8$ TeV
and $\Lumint=20$ fb$^{-1}$ under the assumption that
no significant excess in the overall number of $WW$ events is observed in the data.
$Z^\prime \to WW$ effects at the 8~TeV LHC with a luminosity of $4.8~\text{fb}^{-1}$ have been discussed in \cite{GonzalezFraile:2012fq,Eboli:2011ye,ATLAS:2012mec,Chatrchyan:2013yaa}.\footnote{Recent studies \cite{Curtin:2014zua,Kim:2014eva} claim a small excess compatible with stop production and decay.}
The form of these bounds reflects the fact that the number of background events is below 3, and that the criterion $N_{Z^\prime}<3$ is the crucial one.
For comparison, also the results for 14~TeV and 100 fb$^{-1}$ (shown in Fig.~\ref{fig3}) are included,
together with current limits on $M_2$ derived from
the Drell--Yan process, $pp\to l^+l^-+X$, at 8~TeV, 
and constraints from electroweak data.
Fig.~\ref{fig6} shows that the current limits on
$\phi$ from the EW precision data are stronger than those obtained
from the present LHC data collected from the 8~TeV run, while the
LHC with 14 TeV possesses a high potential to substantially improve the
current bounds on the $Z$-$Z'$ mixing angle.

\subsection{Semileptonic $WW$ decays}

\begin{figure}[htb] \vspace{-3mm}
\begin{center}
{\includegraphics[angle=0, scale=1.1]{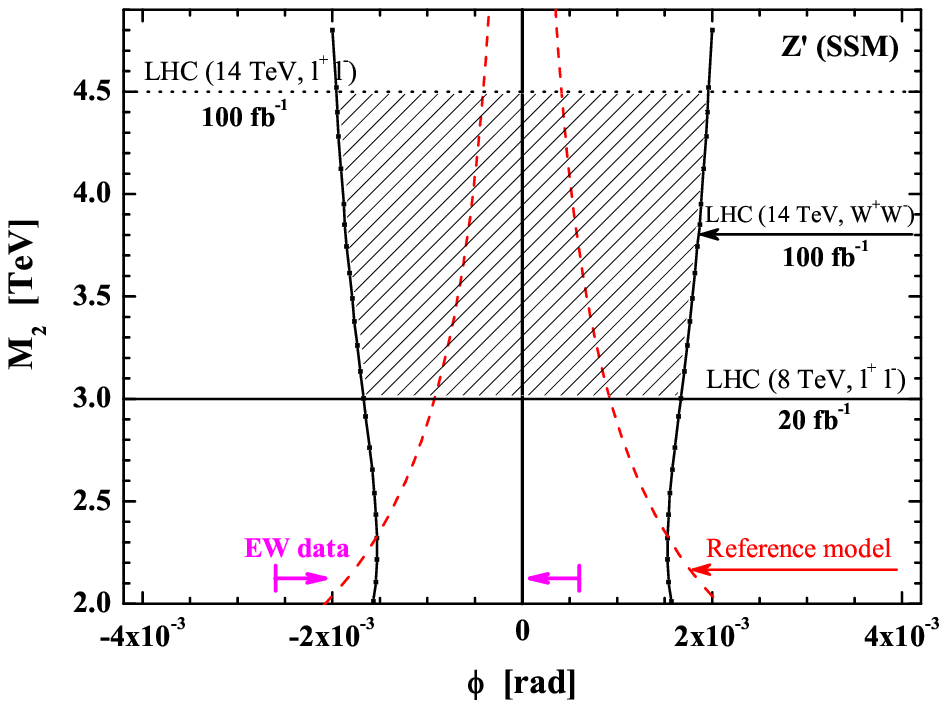}}
\end{center}
\vspace{-3mm}
\caption{Same as in Fig.~3 but obtained from the process $pp\to
WW\to l\nu jj + X$ at the LHC with $\sqrt{s}=14$ TeV and
$\Lumint=100$ fb$^{-1}$. 
The dashed red line shows the predicted relation between $\phi$ and $M_2$, given by Eq.~(\ref{phi0}) for the model of Ref.~\cite{Altarelli:1989ff}.
Note that the scale is different from those of Figs.~\ref{fig3}--\ref{fig6}.}\label{fig7}
\end{figure}

As mentioned above, another process where one can search for a new diboson resonance such as the $Z'$ is represented by the subsequent $WW$ decay into an $l\nu jj$ final state, i.e., a charged lepton (electron or muon), large missing transverse momentum ($E_T^{\rm
miss}$), and at least two jets,
\begin{equation}
pp\to WW+X\to l\nu jj+X. \label{semi}
\end{equation}
An advantage of that process is that it has a higher
crosss section with respect to the pure leptonic final states.
Also, the $l\nu jj$ final state allows the reconstruction of the
invariant mass of the $WW$ system, under certain assumptions for the
neutrino longitudinal momentum from a $W$ boson decay. As a result, a sharper
$Z'$ signal can be obtained. 
On the other hand, this channel has large QCD
backgrounds due to the $Wjj$ production, as well as $Zjj$ with
$Z$ decaying leptonically and one of the leptons being missed.
Also, $t\bar{t}$ production contributes to the background.
However, the large QCD background can be reduced by making use of
the characteristic harder transverse momenta of the charged lepton and
the jets in the $Z'$ signal. A detailed analysis of the QCD
background and the corresponding cuts imposed, resulting in  its
substantial reduction, the estimation on the discovery potential
of the $Z'$ boson in this channel
is given in \cite{Benchekroun:2001je} and, more recently, in \cite{Alves:2009aa}. For the overall
background we refer to Ref.~\cite{Benchekroun:2001je} in our further analysis, in order
to quantify the expected statistical significance as a function of
$M_{Z'}$ for different $Z'$ models. We find that for the
integrated luminosity of 100 $\text{fb}^{-1}$ the semileptonic channel allows
to make further improvement of the current limits on the $Z$-$Z'$
mixing angle as reported in Fig.~\ref{fig7} and summarized in Table~\ref{tab4}.

In Fig.~\ref{fig7} we also show the $\phi$--$M_2$ relation for the specific
`minimal-Higgs model' determined by Eq.~(\ref{phi0}) where ${\cal C}$ is
chosen to be unity \cite{Benchekroun:2001je,Altarelli:1989ff}. 
The possibility of the $Z'$ boson detection in the semileptonic decay
mode of $WW$ at the LHC has been discussed in \cite{Benchekroun:2001je}.
Our numerical results for this model presented in Fig.~\ref{fig7} are
consistent with those given in Ref.~\cite{Benchekroun:2001je}.
However, an improvement with respect to the EW data is for the reference model only possible 
for negative values of $\phi$.
\section{Concluding remarks}
In Table~\ref{tab4}, we collect our limits on the $Z'$ parameters
for the models listed in Section~II. Also shown in
Table~\ref{tab4} are the current limits on various $Z'$ boson
masses from the LEP2 and Tevatron from studies of diboson $W^+W^-$
pair production. The limits on $\phi$ and $M_2$ at the Tevatron
assume that no decay channels into exotic fermions or
superpartners are open to the $Z'$. Otherwise, the limits would be
moderately weaker. LEP2 constrains virtual and $Z$-$Z'$ boson
mixing effects by the angular distribution of $W$ bosons.
Table~\ref{tab4} shows that the limits on $\phi$ from the EW
precision data are generally competitive with and in many cases
stronger than those from the colliders, except for the ILC (1 TeV) and
LHC (14 TeV) that possess high potential to improve substantially
the current bounds on the $Z$-$Z'$ mixing angle. We stress that these
limits are highly complementary.

\begin{table}[htb]
\caption{Reach on the $Z$-$Z^\prime$ mixing angle $\phi$ at
95\% C.L.  in different processes and experiments.}
\begin{center}
\begin{tabular}{|>{\small}c|>{\small}c|>{\small}c|>{\small}c|>{\small}c|>{\small}c|>{\small}c|}
\hline collider, process & $|\phi|$ & $Z'_\chi$ & $Z'_\psi$ & $Z'_\eta$ & $Z'_{\rm SSM}$ & @ $M_{Z^\prime}$\\
\hline LEP2, $e^+e^- \to W^+W^-$ \cite{Andreev:2012zza} & $|\phi|,
10^{-2}$
& 6 & 15 & 50 & 7 & $\geq$ 1 TeV\\
\hline Tevatron, $p\bar{p} \to W^+W^- + X$ \cite{Aaltonen:2010ws}
& $|\phi|, 10^{-2}$
& -- & -- & -- & 2 & 0.4--0.9 TeV\\
\hline electroweak (EW) data \cite{Erler:2009jh} & $|\phi|,
10^{-3}$
& 1.6 & 1.8 & 4.7 & 2.6 &  --\\
\hline ILC (0{.}5~TeV), $e^+e^- \to W^+W^-$ \cite{Andreev:2012cj}&
$|\phi|, 10^{-3}$
& 1{.}5 & 2{.}3 & 1{.}6 & 1{.}2 & $\geq$ 3 TeV\\
\hline ILC (1{.}0~TeV), $e^+e^- \to W^+W^-$ \cite{Andreev:2012cj}&
$|\phi|, 10^{-3}$
& 0{.}4 & 0{.}6 & 0{.}5 & 0{.}3 & $\geq$ 3 TeV\\
\hline LHC (8~TeV), $pp \to W^+W^- \to l\nu l^\prime\nu^\prime$
(this work)& $|\phi|, 10^{-3}$
& -- & -- & -- & 5.2 & 3 TeV\\
\hline LHC (14~TeV), $pp \to W^+W^- \to l\nu jj$ (this work)&
$|\phi|, 10^{-3}$
& 1.7 & 1.2 & 1.1 & 1.7 & 3 TeV\\
\hline LHC (14~TeV), $pp \to W^+W^- \to l\nu l^\prime\nu^\prime$
(this work) & $|\phi|, 10^{-3}$
& 0.4--0.8 & 0.3--0.6 & 0.3--0.6 & 0.5--0.9 & 3--4.5 TeV\\
\hline
\end{tabular}
\end{center}
\label{tab4}
\end{table}

The diboson-channel limit for $Z'_{\rm LR}$ bosons
from the LHC are numerically very similar to those for the $Z'_\chi$ model, only
slightly lower (not shown).

If a new $Z'$ boson exists in the mass range $\sim$ 3--4.5 TeV, its
discovery is possible in the Drell--Yan channel. Moreover, the
detection of the $Z'\to W^+W^-$ mode is eminently possible and
gives valuable information on the $Z$-$Z'$ mixing. It might be the only
mode other than the dileptonic one, $Z'\to l^+l^-$, that is accessible.
Our results demonstrate that it might be possible to detect a new
heavy $Z'$ boson from the totally leptonic or semileptonic
$WW$ channels at the LHC. The LHC at nominal energy and integrated luminosity
provides the best opportunity of studying a new heavy $Z'$ through
its $WW$ decay mode and creates the possibility of measuring (or
constraining) the $Z$-$Z'$ mixing, thus providing insight into the
pattern of symmetry breaking. 

\section*{Acknowledgments}
It is a pleasure to thank F. Staub for pointing out the importance of kinetic mixing.
This research has been partially supported by the Abdus Salam ICTP
(TRIL Programme),  the Collaborative Research Center SFB676/1-2006
of the DFG at the Department of Physics of the University of
Hamburg and the Belarusian Republican Foundation for Fundamental
Research. The work of PO has been supported by the Research Council of Norway.




\begin{thebibliography}{99}

\bibitem{Beringer:2012}
  J.~Beringer {\it et. al.} (Particle Data Group),
  Phys.\ Rev.\  D {\bf 86}, 010001 (2012).

\bibitem{Hewett:1988xc}
  J.~L.~Hewett and T.~G.~Rizzo,
  Phys.\ Rept.\  {\bf 183}, 193 (1989).

\bibitem{Langacker:2008yv}
  P.~Langacker,
  Rev.\ Mod.\ Phys.\  {\bf 81}, 1199-1228 (2009)
  [arXiv:0801.1345 [hep-ph]].

\bibitem{Leike:1998wr}
  A.~Leike,
  Phys.\ Rept.\  {\bf 317}, 143-250 (1999)
  [hep-ph/9805494].

\bibitem{Gulov:2010zq}
  A.~Gulov and V.~Skalozub,
  Int.\ J.\ Mod.\ Phys.\ A {\bf 25}, 5787 (2010)
  [arXiv:1009.2320 [hep-ph]].

\bibitem{Chatrchyan:2012oaa}
  S.~Chatrchyan {\it et al.}  [CMS Collaboration],
  Phys.\ Lett.\ B {\bf 720}, 63 (2013)
  [arXiv:1212.6175 [hep-ex]].

\bibitem{atlas-dilepton}
  G.~Aad {\it et al.}  [ATLAS Collaboration],
  JHEP {\bf 1211}, 138 (2012)
  [arXiv:1209.2535 [hep-ex]].

\bibitem{Aad:2014cka} 
  G.~Aad {\it et al.}  [ATLAS Collaboration],
  arXiv:1405.4123 [hep-ex].

\bibitem{CMS:2013qca} 
  CMS Collaboration [CMS Collaboration],
  CMS-PAS-EXO-12-061.

\bibitem{Erler:2009jh}
  J.~Erler, P.~Langacker, S.~Munir, E.~Rojas,
  JHEP {\bf 0908}, 017 (2009)
  [arXiv:0906.2435 [hep-ph]].

\bibitem{Andreev:2012zza}
  V.~V.~Andreev and A.~A.~Pankov,
  Phys.\ Atom.\ Nucl.\  {\bf 75}, 76 (2012)
  [Yad.\ Fiz.\  {\bf 75}, 67 (2012)].

\bibitem{Aaltonen:2010ws}
  T.~Aaltonen {\it et al.}  [CDF Collaboration],
  Phys.\ Rev.\ Lett.\  {\bf 104}, 241801 (2010)
  [arXiv:1004.4946 [hep-ex]].

\bibitem{ALEPH:2005ab} 
  S.~Schael {\it et al.}  [ALEPH and DELPHI and L3 and OPAL and SLD and LEP Electroweak Working Group and SLD Electroweak Group and SLD Heavy Flavour Group Collaborations],
  Phys.\ Rept.\  {\bf 427}, 257 (2006)
  [hep-ex/0509008].

\bibitem{Godfrey:2013eta} 
  S.~Godfrey and T.~Martin,
  arXiv:1309.1688 [hep-ph].

\bibitem{Dittmar:2003ir} 
  M.~Dittmar, A.~-S.~Nicollerat and A.~Djouadi,
  Phys.\ Lett.\ B {\bf 583}, 111 (2004)
  [hep-ph/0307020].
  
\bibitem{Rizzo:2006nw}
  T.~G.~Rizzo,
    [hep-ph/0610104].

\bibitem{Andreev:2012cj}
  V.~V.~Andreev, G.~Moortgat-Pick, P.~Osland, A.~A.~Pankov and N.~Paver,
  Eur.\ Phys.\ J.\ C {\bf 72}, 2147 (2012)
  [arXiv:1205.0866 [hep-ph]].

\bibitem{Ananthanarayan:2010bt}
 B.~Ananthanarayan, M.~Patra and P.~Poulose,
 JHEP {\bf 1102}, 043 (2011)
 [arXiv:1012.3566 [hep-ph]].
 
\bibitem{Abazov:2010dj}
  V.~M.~Abazov {\it et al.}  [D0 Collaboration],
  Phys.\ Rev.\ Lett.\  {\bf 107}, 011801 (2011)
  [arXiv:1011.6278 [hep-ex]].

\bibitem{Aad:2013wxa}
  G.~Aad {\it et al.}  [ATLAS Collaboration],
  Phys.\ Rev.\ D {\bf 87}, no. 11, 112006 (2013)
  [arXiv:1305.0125 [hep-ex]].

\bibitem{Aad:2012nev}
  G.~Aad {\it et al.}  [ATLAS Collaboration],
  Phys.\ Lett.\ B {\bf 718}, 860 (2013)
  [arXiv:1208.2880 [hep-ex]].
  
\bibitem{UPR-0476T}
  P.~Langacker and M.~-x.~Luo,
  Phys.\ Rev.\ D\ {\bf 45}, 278  (1992).

\bibitem{Holdom:1985ag} 
  B.~Holdom,
  Phys.\ Lett.\ B {\bf 166}, 196 (1986).

\bibitem{Krauss:2012ku} 
  M.~E.~Krauss, B.~O'Leary, W.~Porod and F.~Staub,
  Phys.\ Rev.\ D {\bf 86}, 055017 (2012)
  [arXiv:1206.3513 [hep-ph]].
  
\bibitem{Hirsch:2012kv} 
  M.~Hirsch, W.~Porod, L.~Reichert and F.~Staub,
  Phys.\ Rev.\ D {\bf 86}, 093018 (2012)
  [arXiv:1206.3516 [hep-ph]].

\bibitem{Campbell:1999ah} 
  J.~M.~Campbell and R.~K.~Ellis,
  Phys.\ Rev.\ D {\bf 60}, 113006 (1999)
  [hep-ph/9905386].

\bibitem{Campbell:2011bn} 
  J.~M.~Campbell, R.~K.~Ellis and C.~Williams,
  JHEP {\bf 1107}, 018 (2011)
  [arXiv:1105.0020 [hep-ph]].
 
\bibitem{Agarwal:2010sp}
  N.~Agarwal, V.~Ravindran, V.~K.~Tiwari and A.~Tripathi,
  Phys.\ Rev.\ D {\bf 82}, 036001 (2010)
  [arXiv:1003.5450 [hep-ph]].

\bibitem{Bai:2012zza}
  Y.~-M.~Bai, L.~Guo, X.~-Z.~Li, W.~-G.~Ma and R.~-Y.~Zhang,
  Phys.\ Rev.\ D {\bf 85}, 016008 (2012)
  [arXiv:1112.4894 [hep-ph]].

\bibitem{Pumplin:2002vw}
J.~Pumplin, D.~R.~Stump, J.~Huston, H.~L.~Lai, P.~Nadolsky and
W.~K.~Tung,
  JHEP {\bf 0207}, 012 (2002)
  [arXiv:hep-ph/0201195].
  
\bibitem{Nuss1997}
  E.~Nuss,
  Z.\ Phys.\ C\ {\bf 76}, 701 (1997).

\bibitem{Osland:2008sy}
  P.~Osland, A.~Pankov, N.~Paver, A.~Tsytrinov,
  Phys.\ Rev.\ D\ {\bf 78}, 035008 (2008).

\bibitem{Atlas} ATLAS Collaboration, Reports
No. CERN-LHCC-99-14, CERN-LHCC-99-15.

\bibitem{Gounaris:1992kp} 
  G.~Gounaris, J.~Layssac, G.~Moultaka and F.~M.~Renard,
  Int.\ J.\ Mod.\ Phys.\ A {\bf 8}, 3285 (1993).

\bibitem{Gounaris:2013sva}
 G.~J.~Gounaris and F.~M.~Renard,
 arXiv:1309.3177 [hep-ph].
 
\bibitem{Pankov:1992cy} 
  A.~A.~Pankov and N.~Paver,
  Phys.\ Rev.\ D {\bf 48}, 63 (1993).

\bibitem{Accomando:2013sfa} 
  E.~Accomando, D.~Becciolini, A.~Belyaev, S.~Moretti and C.~Shepherd-Themistocleous,
  JHEP {\bf 1310}, 153 (2013)
  [arXiv:1304.6700 [hep-ph]].
  
\bibitem{Benchekroun:2001je}
  D.~Benchekroun, C.~Driouichi and A.~Hoummada,
  Eur.\ Phys.\ J.\ direct C {\bf 3}, N3 (2001).

\bibitem{delAguila:1987af} 
  F.~del Aguila, L.~Ametller, R.~D.~Field and L.~Garrido,
  Phys.\ Lett.\ B {\bf 201}, 375 (1988).

\bibitem{Agarwal:2010sn}
  N.~Agarwal, V.~Ravindran, V.~K.~Tiwari and A.~Tripathi,
  Phys.\ Lett.\ B {\bf 690}, 390 (2010)
  [arXiv:1003.5445 [hep-ph]].

\bibitem{Cooke_2008}
M.~P.~Cooke, ``WW production cross section measurement and limits
on anomalous trilinear gauge couplings at $\sqrt{s} = 1.96$ TeV'',
Thesis, Rice University, Houston, Texas, 2008, 163 pages.

\bibitem{Alves:2009aa}
  A.~Alves, O.~J.~P.~Eboli, D.~Goncalves, M.~C.~Gonzalez-Garcia and J.~K.~Mizukoshi,
  Phys.\ Rev.\ D {\bf 80}, 073011 (2009)
  [arXiv:0907.2915 [hep-ph]].

\bibitem{Eboli:2011bq}
  O.~J.~P.~Eboli, C.~S.~Fong, J.~Gonzalez-Fraile and M.~C.~Gonzalez-Garcia,
  Phys.\ Rev.\ D {\bf 83}, 095014 (2011)
  [arXiv:1102.3429 [hep-ph]].

\bibitem{GonzalezFraile:2012fq} 
  J.~Gonzalez-Fraile,
  arXiv:1205.5802 [hep-ph].

\bibitem{Eboli:2011ye} 
  O.~J.~P.~Eboli, J.~Gonzalez-Fraile and M.~C.~Gonzalez-Garcia,
  Phys.\ Rev.\ D {\bf 85}, 055019 (2012)
  [arXiv:1112.0316 [hep-ph]].

\bibitem{ATLAS:2012mec} 
  G.~Aad {\it et al.}  [ATLAS Collaboration],
  Phys.\ Rev.\ D {\bf 87}, no. 11, 112001 (2013)
  [Erratum-ibid.\ D {\bf 88}, no. 7, 079906 (2013)]
  [arXiv:1210.2979 [hep-ex]].
  
\bibitem{Chatrchyan:2013yaa} 
  S.~Chatrchyan {\it et al.}  [CMS Collaboration],
  Eur.\ Phys.\ J.\ C {\bf 73}, 2610 (2013)
  [arXiv:1306.1126 [hep-ex]].

\bibitem{Curtin:2014zua} 
  D.~Curtin, P.~Meade and P.~J.~Tien,
  arXiv:1406.0848 [hep-ph].
 
\bibitem{Kim:2014eva} 
  J.~S.~Kim, K.~Rolbiecki, K.~Sakurai and J.~Tattersall,
  arXiv:1406.0858 [hep-ph].

\bibitem{Altarelli:1989ff}
  G.~Altarelli, B.~Mele and M.~Ruiz-Altaba,
  Z.\ Phys.\ C {\bf 45}, 109 (1989)
  [Erratum-ibid.\ C {\bf 47}, 676 (1990)].

\end{thebibliography}
\end{document}